\documentclass[11pt,a4]{article}
\usepackage[latin1]{inputenc}
\usepackage[english]{babel}
\usepackage{amsmath}
\usepackage{amssymb}
\usepackage{fancybox}
\usepackage{geometry}
\usepackage{latexsym}
\usepackage{hyperref}

\begin{document}
\selectlanguage{english}

\newenvironment{eq}{\vskip3ex\begin{center}
 \begin{Sbox}\begin{minipage}{0.9\textwidth}{}}
 {\end{minipage}\end{Sbox}\shadowbox{\TheSbox}\end{center}\vskip2ex}%The vskip is the distance below the box

\begin{eq}
\vskip2ex
\centering \huge The computer BESK and an early attempt \\ to simulate galactic dynamics \\ \ \\ {\normalsize \bf Per Olof Lindblad} \\ \ \\ \small Stockholm Observatory, Stockholm University, 106 91 Stockholm, Sweden\\
\end{eq}

\noindent
The first N-body simulation of interacting galaxies, even producing spiral arms, was performed by Erik Holmberg in Lund (1941), not with a numerical computer, but by his arrangement of movable light-bulbs and photocells to measure the luminosity at each bulb and thereby estimate the gravitational force. A decade later, and with the arrival of the first programable computers, computations of galactic dynamics were performed, which were later transferred into a N-body simulation movie. I present here the background details for this work with a description of the important elements to note in the movie which may be retrieved at \url{http://ttt.astro.su.se/\~po}.\\

\noindent
{\bf The computer BESK}\\

\noindent
In the late 1940ies the Swedish Government set up a State board given the task to construct a modern, new generation computer to meet the country's computing demands. In 1953 the computer BESK\footnote{\url{http://en.wikipedia.org/wiki/BESK}.} (Binary Electronic Sequence Calculator) was completed and ready to start. It then had a Williams Tube based, later changed to ferrite-core, direct access memory containing a maximum of 2048 numbers or instructions. An addition or subtraction was done in 41 microseconds. The operation times for multiplication and division were 266 and 512 microseconds respectively. Input and output were performed via punch tape. \\

\noindent
The computer was used already from the early stages by Ingrid Torg\aa rd in Lund and Alexander Ollongren in Leiden (1962) for computations of three-dimensional galactic orbits, and by Per Olof Lindblad in Stockholm (1960) for two-dimensional galactic N-body simulations. One of these early simulations is shown in the accompanying movie, which will illustrate some aspects of galactic dynamics.\\

\noindent
{\bf The N-body computing program}\\

\noindent
From the very beginning it occurred that an interesting task for this computer would be to simulate a galaxy by an N-body problem. The memory size of the computer then puts a strict upper limit to the number of bodies. The fourth order Runge-Kutta integration method used, requires a fair amount of memory space. As far as only the direct access memory was used, and provided the bodies were placed in a two-dimensional and strictly bi-symmetric configuration, the maximum number of bodies was restricted to N = 116. \\

\noindent
This restricted number of bodies, confined to a plane, cannot realistically simulate the mass distribution in an entire galaxy. Thus, the main part of the gravitational force in the plane of the galaxy was represented by a fixed rotationally symmetric central force, while the free mass points represented a thin disk. For the fixed central force, the force derived by Marten Schmidt (1956) for the Galactic System was chosen, and the mass of each point was chosen to be 64 million solar masses. Their total mass of the free bodies then would be of the order of 4\% of the mass of the galaxy. To avoid violent encounters between the mass points, a sphere with `soft gravity' surrounding each mass point had to be introduced. The radii of these spheres decide the rate of the ongoing increase of velocity dispersion occurring throughout the computation.\\

\noindent
{\bf The galactic model}\\

\noindent
In the case illustrated in the movie, the mass points were placed in three circular rings with radii of 2, 4 and 6 kpc and given initial circular rotational velocities. In order to throw this system out of balance, the mass points in the middle ring were shifted slightly towards two bisymmetric positions along the ring as indicated in the starting sequence of the movie. Then the computation is started, keeping a full integration step of 8 million years, each step requiring 7.5 minutes of computing time. The clock in the upper right corner of the frame indicates the number of years that have elapsed since the start. \\

\noindent
{\bf To note in the movie ( \url{http://ttt.astro.su.se/\~po} ) }\\

\noindent
The immediate reaction on the uneven mass distribution in the middle ring is that the points trailing the density maxima are accelerated and move outwards, while the points leading the maxima are decelerated and move inwards, giving the ring a slight elongated shape. In fact two interfering elongated waves are exited, one moving slower than circular motion and hence with density maxima at the ends of the major axis and one moving faster than circular motion and hence with density maxima at the minor axis. With these two waves superimposed, the result is that the middle ring for a while oscillates between elongated shape with even mass distribution, and almost circular shape with bipolar mass distribution, i,e, it exerts a bar-like perturbation on its surroundings. The outer ring lies outside the co-rotation resonance of this bar-like perturbation of the middle ring. This causes the outer ring to strive for an elongated shape with major axis at right angles to that of the bar and with density maxima at the ends of its minor axis. By mutual interactions, the amplitudes increase, and after about 350 million years from the start the two rings coalesce, the rings break up and from 400 million years form a pair of leading spiral arms. \\

\noindent
These leading spiral arms are not stable. The leading parts of the arms are pulled back by the gravitational attraction of the arms, which makes them fall inwards and then shoot forwards in what Donald Lynden-Bell has called a `donkey behaviour'. The trailing parts of the arms, on the other hand, are pulled forward, move outwards and lag behind, again a `donkey behaviour'. In this way, the middle and the outer arms change place, and after 550 million years now form a pair of trailing arms. It should be noted that this change from leading to trailing arms is not due to differential rotation but to self-gravity within the arms that turn them inside out. (This is more clearly seen in the printed version of the simulation, where point masses from different rings are marked with different symbols.) \\

\noindent
The trailing arms are subsequently drawn out by differential rotation, until they by 750 years vanish with increasing velocity dispersion. However, now the inner ring in turn has acquired an elongated shape acting as a bar, which again creates a trailing spiral structure formed mainly by point masses from the middle ring, 900 million years from the start, and after a total computer time of about 15 hours. \\

\noindent
The present movie was prepared during a stay at the Leiden Observatory in the spring of 1961 for the symposium Interstellar matter in galaxies (ed. L. Woltjer, W.A. Benjamin, Inc., p. 222, 1962) held in Princeton in April 1961. Today of course, there have appeared innumerable, infinitely more sophisticated, simulations of galaxies.\\

\noindent
{\bf References}
\begin{itemize}
\item Holmberg, E., 1941, ApJ, 94, 385 \\\url{http://adsabs.harvard.edu/abs/1941ApJ....94..385H}
\item Lindblad, P.O., 1960, Stockholm Obs. Ann. 21, No. 4 \\\url{http://adsabs.harvard.edu/abs/1960StoAn..21....4L}\\\url{http://ttt.astro.su.se/\~po}
\item Ollongren, A., 1962, Bull. Astron. Inst. Netherlands 16, 241\\\url{http://adsabs.harvard.edu/abs/1962BAN....16..241O}
\item Schmidt, M., 1956, Bull. Astron. Inst. Netherlands 13, 15 \\\url{http://adsabs.harvard.edu/abs/1956BAN....13...15S}
\end{itemize}

\end{document}